%Paper: gr-qc/9504017
%From: "P. Suntharothok-Priesmeyer" <psp@howdy.wustl.edu>
%Date: Wed, 12 Apr 1995 08:13:44 -0500

\magnification=1200
\voffset -.3truecm \vsize 25truecm
\overfullrule=0pt
\tolerance=1000
\def\br{\hfil\break}
\def\newpage{\vfil\eject}
\headline={\ifnum\pageno = 1\hss\tenrm WUGRAV-95-5\else\hss\fi}
\footline={\hss}
\null\vskip 3.4truecm
\centerline{\bf STABLE CLOCKS AND GENERAL RELATIVITY}
\bigskip
\centerline{Clifford M. Will}
\centerline{McDonnell Center for the Space Sciences, Department of Physics}
\centerline{Washington University, St. Louis, Missouri 63130, U.S.A.}
\vskip 9truecm
\centerline{ABSTRACT}
\medskip
We survey the role of stable clocks in general relativity.  Clock
comparisons have provided important tests of the Einstein Equivalence
Principle, which underlies metric gravity.  These include tests of
the isotropy of clock comparisons (verification of local Lorentz
invariance) and tests of the homogeneity of clock comparisons
(verification of local position invariance).  Comparisons of atomic
clocks with gravitational clocks test the Strong Equivalence Principle
by bounding cosmological variations in Newton's constant.  Stable
clocks also play a role in the search for gravitational radiation:
comparision of atomic clocks with the binary pulsar's orbital clock
has verified gravitational-wave damping, and phase-sensitive detection
of waves from inspiralling compact binaries using laser interferometric
gravitational observatories will facilitate extraction of useful source
information from the data.  Stable clocks together with general
relativity have found important practical applications in navigational
systems such as GPS.
\newpage
\baselineskip 5mm
\leftline{\bf 1.\quad Introduction}
\smallskip

Stable clocks have long played an important role in the field of
general relativity.  In addition to the use of the notion of the
``ideal clock'' in thought experiments that have become part of the
conceptual foundation of special and general relativity, stable
clocks realized in the laboratory have been employed extensively
in experimental tests of relativity.  They have been used as direct
tools to measure or test relativistic effects, such as in tests
of the gravitational redshift.  But they have also been used as
indirect, supporting tools, such as in the measurement of
gravitational-radiation damping of the orbit of the binary pulsar.
In fact, from one point of view, most experimental tests of special
or general relativity amount to nothing more than comparisons of
stable clocks.  Finally, stable clocks, together with general
relativity, have also found recent practical use in high-precision
navigational systems, such as the Global Positioning System (GPS).

In this paper, we survey some of the varied uses of stable clocks
in general relativity.  In Section~II we describe the Einstein
Equivalence Principle, which is the foundation for the idea that
spacetime is curved (and thereby a foundation for general relativity),
and discuss the clock experiments that support it.  Many of the
experiments involve various kinds of intercomparisons of
clocks --- comparing identical clocks at different locations;
comparing different clocks at the same location but with varying
orientation; or comparing different clocks at a single location
that moves relative to distant matter.  In Section~III we discuss
the Strong Equivalence Principle and intercomparisons between
atomic and gravitational clocks.  Section~IV discusses the use
of stable clocks in the verification of and search for gravitational
radiation, and Section~V describes the use of stable clocks
and general relativity in GPS.
\bigskip
\leftline{\bf 2.\quad Stable Clocks and Metric Gravity }
\medskip
\noindent
{\bf 2.1 The Einstein Equivalence Principle}
\smallskip
The Einstein Equivalence Principle is the foundation for all metric
theories of gravity, such as general relativity, Brans-Dicke theory
and many others (for a review of topics and concepts discussed
in this paper see Ref.~1).  It states, roughly, that all test bodies
fall in a gravitational field with the same acceleration
(Weak Equivalence Principle), and that in local, freely falling
or inertial frames, the outcomes of non-gravitational experiments
are independent of the velocity of the frame (Local Lorentz Invariance)
and the location of the frame (Local Position Invariance).  A consequence
of this principle is that the non-gravitational interactions must
couple {\it only} to the symmetric spacetime metric $g_{\mu\nu}$,
which locally has the Minkowski form $\eta_{\mu\nu}$ of special relativity.
Because of this local interaction only with $\eta_{\mu\nu}$,
local nongravitational physics is immune from the influence of distant
matter, apart from tidal effects.  Local physics is Lorentz invariant
(because $\eta_{\mu\nu}$ is), and position invariant (because
$\eta_{\mu\nu}$ is constant in space and time).

How could violations of EEP arise?  From the viewpoint of field theory,
violations of EEP would generically be caused by other long-range
fields additional to $g_{\mu\nu}$ which also couple to matter,
such as scalar, vector and tensor fields.  Such theories are called
non-metric theories.  A simple example of a non-metric theory is one
in which the matter action for charged particles is given by
$$
\eqalignno{I=
&-\sum_a m_a \int ( g_{\mu\nu} v_a^\mu v_a^\nu )^{1/2} dt
  + \sum_a e_a \int A_\mu (x_a^\nu ) v_a^\mu dt\cr
&- (16\pi)^{-1} \int \sqrt{-h} h^{\mu\alpha} h^{\nu\beta}
   F_{\mu\nu} F_{\alpha\beta} d^4 x \,, &(1)\cr}
$$
where $m_a$, $e_a$, $x_a^\nu$, and $v_a^\nu =dx_a^\nu /dt$ are the
mass, charge, world-line and ordinary velocity, respectively, of the
$a$-th body, $A_\mu$ and $F_{\mu\nu}$ are the electromagnetic vector
potential and Maxwell field, $g_{\mu\nu}$ is the metric,
$h_{\mu\nu}$ is a second, second-rank tensor field, and repeated
Greek indices imply summation over four spacetime values.
Locally one can always find coordinates (local inertial frame)
in which $g_{\mu\nu} \to \eta_{\mu\nu}$, but in general
$h_{\mu\nu} \not\to \eta_{\mu\nu}$,
instead, $h_{\mu\nu} \to h_{\mu\nu}^0$, where $h_{\mu\nu}^0$
is a tensor whose values are determined by the cosmological
or nearby matter distribution.  In the rest frame of the distant
matter distribution, $h_{\mu\nu}^0$ will have specific values,
and there is no reason {\it a priori} why those should correspond
to the Minkowski metric (unless $h_{\mu\nu}$ were identical to
$g_{\mu\nu}$ in the first place, in which case one would
have a metric theory).  The values of $h_{\mu\nu}^0$ could also
vary with the location of the local frame in space or time relative
to the distant matter.  This can lead to violations of Lorentz
invariance or position invariance in the local physics of
electromagnetic systems.

A number of explicit theoretical frameworks have been developed
to treat a broad range of non-metric theories, of which this was
just one example.  They include the $TH\epsilon\mu$ framework
of Lightman and Lee,$^{2)}$ the $\chi - g$ framework of Ni,$^{3)}$
the $c^2$ framework of Haugan and coworkers,$^{4,5)}$ and the
extended $TH\epsilon\mu$ framework of Vucetich and colleagues.$^{6)}$

\medskip
\noindent
{\bf 2.2 Local Lorentz Invariance}
\smallskip
Tests of Local Lorentz Invariance are most profitably discussed
using the $c^2$ Framework.  This is a special case of the
$TH\epsilon\mu$ formalism, adapted to situations in which one can
ignore the variation with space and time of the external fields that
couple to matter, and instead focus on their dependence on the
velocity of the local frame.  It assumes a class of non-metric
theories in which the matter part of the action of Eq.~(1) can
be put into the local special relativistic form, using units
in which the limiting speed of neutral test bodies is unity,
and in which the sole effect of any non-metric fields coupling
to electrodynamics is to alter the effective speed of light.
The result is the action
$$
I = -\sum_a m_a \int \sqrt{1-v_a^2} dt
    + \sum_a e_a \int A_\mu v_a^\mu dt
    + (8\pi)^{-1} \int (E^2-c^2 B^2 ) d^4x \,,\eqno{(2)}$$
where $E$ and $B$ are the usual electric and magnetic fields defined
using components of $F_{\mu\nu}$.  Because the action is explicitly
non-Lorentz invariant if $c^2 \ne 1$, it must be defined in a
preferred universal rest frame (presumably that of the 3K microwave
background); in this frame, the value of $c^2$ is then determined
by the cosmological values of the non-metric field.
Even if the non-metric field coupling to electrodynamics is a tensor
field, the homogeneity and isotropy of the background cosmology
in the preferred frame is likely to collapse its effects to that
of the single parameter $c^2$.  Because this action violates Lorentz
invariance, systems moving through the universe will exhibit explicit
effects dependent upon the velocity of motion.  Detailed calculations
of a variety of experimental situations show that those effects depend
on the magnitude of the velocity through the preferred frame
($\sim 300$~km/sec), and on the parameter $\delta \equiv c^{-2} -1$.
Those effects also depend in general on the internal structure or dynamics
of the system (clock) under study.  In any metric theory or theory with
local Lorentz invariance, $\delta =0$, and no such effects occur,
regardless of the internal structure of the system.

One can then set observable upper bounds on $\delta$ using a variety
of experiments.  Modest bounds on $\delta$ can be set by the ``standard''
tests of special relativity, such as the Michelson-Morley experiment
and its descendents,$^{7,8)}$ or the Brillet-Hall$^{9)}$ interferometry
experiment.  In these examples the two clocks are the two arms of the
interferometers, and the comparison is of their rates (round-trip time
of flight of light) as the arms' orientation varies relative to the
velocity of the Earth through the universe.  Other tests of special
relativity involve comparison of identical atomic clocks separated in
space, as the orientation of their baseline varies; communication
between the clocks is by light propagation.  These include a test of
time-dilation using radionuclides on centrifuges,$^{10)}$ tests of the
relativistic Doppler shift formula using two-photon absorption
(TPA),$^{11)}$ and a test of the isotropy of the speed of light using
one-way propagation of light between hydrogen maser atomic clocks at
the Jet Propulsion Laboratory (JPL).$^{12)}$

Very stringent bounds $| \delta | <10^{-21}$ have been set by
``mass isotropy'' experiments of a kind pioneered by Hughes
and Drever.$^{13,14)}~~$   The idea is simple:  in a frame moving
relative to the preferred frame, the non-Lorentz-invariant
electromagnetic action of Eq.~(2) becomes anisotropic,
dependent on the direction of the velocity $\vec V$.
Those anisotropies then are reflected in the energy levels
of electromagnetically bound atoms and nuclei (for nuclei,
we consider only the electromagnetic contributions).
For example, the three sublevels of an $l=1$ atomic wavefunction
in an otherwise spherically symmetric atom can be split in energy,
because the anisotropic perturbations arising from the electromagnetic
action affect the energy of each substate differently.  One can study
such energy anisotropies by first splitting the sublevels slightly
using a magnetic field, and then monitoring the resulting Zeeman
splitting as the rotation of the Earth causes the laboratory
$\vec B$-field (and hence the quantization axis) to rotate relative
to $\vec V$, causing the relative energies of the sublevels to vary
among themselves diurnally.  Using nuclear magnetic resonance techniques,
the original Hughes-Drever experiments placed a bound of about
$10^{-16}$~eV on such variations.  This is about $10^{-22}$
of the electromagnetic energy of the nuclei used.  Since the magnitude
of the predicted effect depends on the product $V^2 \delta$, and
$V^2 \approx 10^{-6}$, one obtains the bound $| \delta |<10^{-16}$.
Energy anisotropy experiments were improved dramatically in the 1980s
using laser-cooled trapped atoms and ions.$^{15-17)}~~$  This technique
made it possible to reduce the broadening of resonance lines caused
by collisions, leading to improved bounds on $\delta$ shown in
Figure~1 (experiments labelled NIST, U. Washington and Harvard,
respectively).
\newpage
\null\vskip 12truecm
\baselineskip=4mm
\leftline{\hskip 1truecm{Figure 1. Selected tests of Local Lorentz Invariance
showing bounds on the}}
\leftline{\hskip 1truecm{parameter $\delta$, which measures
the degree of violation of Lorentz invariance}}
\leftline{\hskip 1truecm{in electromagnetism.  Michelson-Morley, Joos,
and Brillet-Hall experiments}}
\leftline{\hskip 1truecm{test isotropy of the round-trip speed of light
in interferometers, the later}}
\leftline{\hskip 1truecm{experiment using laser technology.
Two-photon absorption (TPA) and JPL}}
\leftline{\hskip 1truecm{experiments test isotropy of the speed of light
in one-way configurations.}}
\leftline{\hskip 1truecm{The remaining four experiments test isotropy
of nuclear energy levels.}}
\leftline{\hskip 1truecm{Limits assume the speed of the Earth is 300~km/s
relative to the mean}}
\leftline{\hskip 1truecm{rest frame of the cosmic microwave background.}}
\baselineskip 5mm
\bigskip
\medskip
\noindent
{\bf 2.3 Local Position Invariance}
\smallskip

Violations of EEP can also lead to time- and position-dependence
of local physics.  In the model example of Eq.~(1), the values
of $h_{\mu\nu}^0$ imposed by cosmology or by nearby matter could vary,
resulting, for example, in variations of the effective fine-structure
constant, or of the relative rates of atomic clocks.  For example,
in the quantum dynamics of an atomic clock based on the hyperfine
structure of hydrogen (hydrogen maser clock), the components of
$h_{\mu\nu}^0$ in Eq.~(1) will play a different role than they
would say, in the dynamics of a clock based on the resonant
frequency of a microwave cavity, because the role of electromagnetism
is different in the two cases.  If one type of clock is chosen as
a reference standard, then the relative rates of other types of clocks
measured against the standard in local freely falling frames will
generally depend on the location of the frame in space or time.
It is straightforward to show from this that the frequency shift
$\Delta f$ in the comparison of two identical clocks at different
heights in a gravitational potential $U$ will be given by
$\Delta f/f = (1+\alpha) \Delta U/c^2$, where $\alpha$ generally
depends on the type of clock being used $^{1)}$.  If EEP is satisfied,
$\alpha=0$ for {\it all} clocks, and one has the standard
gravitational redshift prediction of Einstein, indeed of all metric
theories of gravity.  In the first gravitational redshift experiment,
the 1960-1965 Pound-Rebka-Snider experiments,$^{18,19)}$ two identical
clocks (gamma-ray emitting iron nuclei) at different heights were
intercompared, leading to a one percent test ($|\alpha_{\rm Fe} |<10^{-2}$).
The best bound to date, $|\alpha_{\rm H-maser}| < 2\times 10^{-4}$,
comes from a 1976 gravitational redshift experiment using a Hydrogen
maser clock launched on a Scout rocket to an altitude of 10,000~km,
and compared with an identical clock on the ground.$^{20)}$

Another class of experiments compares two different clocks side by side,
as the Earth's orbital motion and rotation moves the laboratory in
and out of the Sun's gravitational field, causing annual and diurnal
variations in $U$.  In one experiment, a hydrogen maser clock
(actually a pair of masers) was compared with a set of oscillator clocks
stabilized by superconducting microwave cavities (called SCSO clocks),
resulting in the bound
$|\alpha_{\rm H-Maser}-\alpha_{\rm SCSO}|<10^{-2}$.$^{21)}~~$
A recent comparison of a cesium standard against a magnesium
fine-structure standard over a 430-day period placed a bound on an
annual relative variation at the level
$|\alpha_{\rm Cs}-\alpha_{\rm Mg}|<7 \times 10^{-4}$.$^{22)}~~$  A
comparison involving a hydrogen maser and a trapped mercury ion
standard is also planned.$^{23)}$

The effective fundamental non-gravitational constants of physics
can vary with cosmological time if EEP is violated.  Bounds on such
variations have been obtained from a variety of geological, laboratory,
and astronomical observations.  The best bound, especially for the
fine-structure constant, comes from the Oklo natural fission reactor
in Gabon, Africa, where the natural occurrence of sustained fission
about two billion years ago permits a comparison of the values of
various constants affecting nuclear reactions then with the current
values.  For the fine structure constant, the bound is better that one
part in $10^5$ per 20 billion years.$^{24)}~~$  Clock comparison
experiments have yielded a bound of $7 \times 10^{-4}$ per 20 billion
years.$^{23)}$

\medskip
\noindent
{\bf 2.4 The Weak Equivalence Principle}
\smallskip

The third element of EEP is the Weak Equivalence Principle,
which states that bodies fall with the same acceleration,
independently of their internal structure or composition.
Although tests of WEP do not generally involve stable clocks,
they are worth mentioning here, if only because of the important
role played by the Rencontres de Moriond during the period
1987-92 as an annual meeting for investigators working on
``fifth-force'' experiments.  Many of those experiments were also
tests of WEP.  The current bounds on the fractional difference
in acceleration in the solar or terrestrial gravitational fields
between bodies of different composition are between $10^{-11}$
and $10^{-12}$.$^{25-28)}~~$  Further improvements in fifth-force
experiments are likely to yield bounds tighter by a few orders of
magnitude; a satellite test of the equivalence principle has also
been proposed that could yield a test at the $10^{-17}$ level.

\bigskip
\noindent
{\bf 3.\quad Stable Clocks and the Strong Equivalence Principle}
\medskip
\noindent
{\bf 3.1 The Strong Equivalence Principle}
\smallskip

The Strong Equivalence Principle (SEP) is a generalization of EEP
which states that in local ``freely-falling'' frames that are large
enough to include gravitating systems (such as planets, stars,
a Cavendish experiment, a binary system, {\it etc.}), yet that
are small enough to ignore tidal gravitational effects from
surrounding matter, local {\it gravitational} physics should
be independent of the velocity of the frame and of its location
in space and time.  Also {\it all} bodies, including those bound
by their own self-gravity, should fall with the same acceleration.
General relativity satisfies SEP, whereas most other
metric theories do not ({\it eg.} the Brans-Dicke theory).

It is straightforward to see how a gravitational theory could violate
SEP.$^{29)}~~$  Most alternative metric theories of gravity introduce
auxiliary fields which couple to the metric (in a metric theory they
can't couple to matter), and the boundary values of these auxiliary
fields determined either by cosmology or by distant matter can act
back on the local gravitational dynamics.  The effects can include
variations in time and space of the locally measured effective
Newtonian gravitational constant $G$ (preferred-location effects),
as well as effects resulting from the motion of the frame relative
to a preferred cosmic reference frame (preferred-frame effects).
Theories with auxiliary scalar fields, such as the Brans-Dicke theory
and its generalizations, generically cause temporal and spatial
variations in $G$, but respect the ``Lorentz invariance'' of gravity,
{\it i.e.} produce no preferred-frame effects.  The reason is that
a scalar field is invariant under boosts.  On the other hand,
theories with auxiliary vector or tensor fields can cause
preferred-frame effects, in addition to temporal and spatial
variations in local gravitational physics.  For example, a timelike,
long-range vector field singles out a preferred universal rest frame,
one in which the field has no spatial components; if this field
is generated by a cosmic distribution of matter, it is natural
to assume that this special frame is the mean rest frame of that matter.

General relativity embodies SEP because it contains only one
gravitational field $g_{\mu\nu}$.  Far from a local gravitating system,
this metric can always be transformed to the Min\-kow\-ski form
$\eta_{\mu\nu}$ (modulo tidal effects of distant matter and
$1/r$ contributions from the far field of the local system),
a form that is constant and Lorentz invariant, and thus that
does not lead to preferred-frame or preferred-location effects.

\medskip
\noindent
{\bf 3.2 Cosmological Variation of Newton's Constant}
\smallskip

In metric theories of gravity that violate SEP, $G$ may also vary with
the evolution of the structure of the universe, via the cosmologically
imposed boundary values on the auxiliary fields. In fact, a cosmic
variation in $G$ was a consideration that partly motivated Dicke
to develop the scalar-tensor theory.  Varying $G$ is common in
the various generalized scalar-tensor theories developed recently
for inflationary cosmology.  On the other hand, in a wide class
of such theories, the variations can be large in the early universe
(leading to the desired cosmological consequences), but damp out as
the present epoch is approached.  In many such theories, general
relativity is a natural ``attractor'' to which the cosmic evolution
naturally leads the theory as the conditions of the present universe
are reached.$^{30)}$

Such a variation of $G$ can be tested by comparing a gravitational
clock, such as a planetary orbit, whose period is governed by $G$,
with a stable atomic clock, whose period is governed by atomic
constants.  The best current observational bound,\br
$| \dot G /G | < 4 \times 10^{-12}\,{\rm yr}^{-1}$,
comes from long-term observations of the orbit of Mars
via Viking ranging data $^{31,32)}$.  A similar bound,
$| \dot G /G | < 3 \times 10^{-11}\, {\rm yr}^{-1}$,
comes from timing of the binary pulsar PSR 1855+09.$^{33)}$

\bigskip
\noindent
{\bf 4.\quad Stable Clocks and Gravitational Radiation}
\medskip
\noindent
{\bf 4.1 The Binary Pulsar}
\smallskip
In 1974, the discovery of the binary pulsar PSR 1913+16 by Hulse
and Taylor$^{34)}$ provided a new laboratory for studying general
relativistic effects, and a new arena for the use of stable clocks.
The system consists of a 59 ms period pulsar in an eight-hour orbit
with a companion that has not been seen directly, but that,
on evolutionary grounds, is generally believed to be another
neutron star (actually a dead pulsar).  The basic measurement
technique involves comparing the pulsar ``clock'' with Earth-based
atomic clocks, specifically by comparing the phases of arriving
radio pulses with those of local oscillators.  The unexpected
stability of the pulsar clock, the cleanliness of the orbit,
and the use of stable atomic clocks and time-transfer using GPS,
allowed radio astronomers to determine the orbital and other
parameters of the system to extraordinary accuracy.
Furthermore, the system is highly relativistic
($v_{\rm orbit} /c \approx 10^{-3}$).  Observation of the relativistic
periastron advance ($4^{\rm o}.22662 \pm 0^{\rm o}.00001 \,{\rm yr}^{-1}$)
and of the effects on pulse arrival times of the gravitational redshift
caused by the companion's gravitational field and of the special
relativistic time dilation caused by the pulsar's orbital motion
(0.05\% accuracy) have been used, assuming that general relativity
is correct, to constrain the nature of the system.  In general relativity,
these two effects depend in a known way on measured orbital parameters
and on the unknown masses $m_p$ and $m_c$ of the pulsar and companion
(assuming that the companion is sufficiently compact that tidal and
rotational distortion effects can be ignored), and consequently
the two masses may be calculated with these two pieces of data,
with the result $m_p = 1.4410 \pm 0.0007 \,M_\odot$
and $m_c = 1.3874 \pm 0.0007 \, M_\odot$.$^{35)}$

A second clock comparison, that of the orbital ``clock'' against
atomic time, gave the first evidence for the effects of gravitational
radiation damping.  General relativity provides a formula, which is
a generalization of one first derived by Einstein in 1916,
known as the quadrupole formula, which determines the loss of energy
and the consequent orbital damping due to gravitational-wave emission
from binary systems such as this.  The result is a decrease in the
orbital period.  Using the measured orbital elements and the two
masses, one can obtain the predicted rate
$dP/dt = -2.40243 \pm 0.00005 \times 10^{-12}$.
The observations are now better than 0.3 percent in accuracy, and it
is necessary to take into account a small effect due to the relative
acceleration between the binary pulsar and the solar system caused by
galactic rotation.$^{36)}~~$  With this effect subtracted, the result is
$dP/dt_{\rm observed} = -(2.410 \pm 0.009) \times 10^{-12}$, agreeing
completely with the prediction.$^{35)}$

Several new binary pulsars have been discovered in the past few years.
Two of these, PSR 1534+12 in our galaxy$^{37)}$ and PSR 2127+11C in the
globular cluster M15,$^{38)}$ are particularly promising as relativity
laboratories.  Combined observations of these systems may yield an even
more accurate determination of $dP/dt$ than did PSR 1913+16 alone.

\medskip
\noindent
{\bf 4.2 Search for Gravitational Waves}
\smallskip
Clock comparisons also play a role in the direct search for
gravitational radiation.  Observations of millisecond pulsars such as
PSR 1937+21 have shown some to be at least as stable as the ensemble
of the world's atomic clocks; the residual timing noise in the comparison
between pulsars and clocks cannot be allocated conclusively to one or the
other.  That residual noise sets a significant upper bound on a
stochastic background of gravitational waves with wavelengths on the
order of light years whose effect would be to cause a fluctuating relative
gravitational redshift between pulsar and Earth times.$^{39,33)}~~$
The resulting bound has ruled out a substantial region of parameter
space for cosmological scenarios involving cosmic strings.

The detection and study of gravitational radiation using large-scale
laser-\-inter\-fero\-metric observatories, such as the U.S. LIGO or the
European VIRGO project, will also rely heavily on clock comparisons,
although not in a manner that will challenge the stability of atomic
standards.$^{40)}~~$  In this case, the broad-band laser systems
will have the ability to track the phase of the incoming GW signal.
For an inspiralling binary system of compact objects (neutron stars
or black holes), one of the most promising detectable sources,
the GW phase evolves non-linearly in time because of
gravitational-radiation damping of the orbit.  That evolution depends
on the physical parameters of the system, such as the masses and spins
of the bodies, and on general relativity.  Using a matched filtering
of theoretical waveform templates against the outputs of the detectors,
it will be possible, because of the intrinsic high precision of phase
comparisons, to determine many of the system parameters to high
accuracy.$^{41)}~~$  However, to achieve that accuracy requires
knowing the general relativistic prediction for the evolution
of the orbital phase of the inspiralling binary to high accuracy,
which translates into a knowledge of gravitational-wave damping
to many orders of approximation beyond the normal quadrupole formula
of standard textbooks.  Written as a formula for the rate of energy loss,
it has the general form
$$
{{dE} \over {dt}}={1 \over 5} \biggl ( {d^3 \over {dt^3}}M^{ij} \biggr )^2
\biggl ( 1+O(\epsilon)+O(\epsilon^{3/2})+O(\epsilon^2)+ \dots \biggr )
\,, \eqno{(3)}
$$
where $M^{ij}$ is the trace-free mass quadrupole moment of the system,
and $\epsilon \sim v^2 \sim m/r$.  Using post-Newtonian and
post-Minkowskian techniques, several workers have succeeded in
deriving this formula for general binary systems through second
post-Newtonian order ($O(\epsilon^2)$), including the effects
of spin,$^{42-44)}$ and calculations to even higher orders are
in progress.  For the special case of a test body inspiralling
onto a black hole, perturbation methods have yielded corrections
to Eq.~(3) through $O(\epsilon^4 \ln \epsilon)$.$^{45,46)}~~$
It is ironic that, for high-precision atomic frequency work,
high-order approximations to solutions of the Schr\"odinger equation
must be used to compare theory with experiment, while here,
it is general relativity that must be approximated to very high-order.

\bigskip
\noindent
{\bf 5.\quad Stable Clocks, General Relativity and GPS}
\smallskip
Recently, stable clocks together with general relativity have begun to
find applications in practical everyday life, through the Global
Positioning System.  This navigation system, based on a constellation
of 24 satellites carrying atomic-clocks, uses precise time transfer
to provide accurate absolute positioning anywhere on Earth to 30~meters,
and differential or relative positioning to the level of centimeters
(a Russian system called GLONASS has similar capabilities).  It relies
on clocks that are stable, run at the same or well calibrated rates,
and are synchronized.  However, the difference in rate between GPS
satellite clocks and ground clocks caused by the gravitational
redshift and time dilation is around 38,000 ns per day.  Consequently,
general relativity must be taken into account in order to achieve the
100 ns time transfer accuracy required for 30 m positioning.  In addition,
the kinematical Sagnac effect must be taken into account in order to
have a consistent clock synchronization scheme on the rotating Earth
(for a discussion of relativity in GPS, see.$^{47)}$)~~
GPS is a classic example of the unexpected and unintended benefits of
basic research --- general relativity early in the century, and masers
and atomic beams in the 1950s.$^{48)}$

\bigskip
\noindent
{\it Acknowledgments.}  This work was supported in part by the National
Science Foundation under Grant No. PHY 92-22902 and by the National
Aeronautics and Space Administration under Grant No. NAGW 3874.

\bigskip
\noindent
REFERENCES
\medskip
\baselineskip=4mm
\item{1.}
C. M. Will, {\it Theory and Experiment in Gravitational
Physics. Revised Edition}, (Cambridge: Cambridge University Press,
1993).
\item{2.}
A. P. Lightman, and D. L. Lee, {\it Phys. Rev. D} {\bf 8}, 364
(1973).
\item{3.}
W.-T. Ni, {\it Phys. Rev. Lett.} {\bf 38}, 301 (1977).
\item{4.}
M. P. Haugan and C. M. Will, {\it Physics Today} {\bf 40}, 69 (May) (1987).
\item{5.}
M. D. Gabriel and M. P. Haugan, {\it Phys. Rev. D} {\bf 41}, 2943 (1990).
\item{6.}
J. E. Horvath, E. A. Logiudice, C. Reveros, and H. Vucetich,
{\it Phys. Rev. D} {\bf 38}, 1754 (1988).
\item{7.}
D. C. Miller, {\it Rev. Mod. Phys.} {\bf 5}, 203 (1933).
\item{8.}
R. S. Shankland, S. W. McCuskey, F. C. Leone, and G. Kuerti,
{\it Rev. Mod. Phys.} {\bf 27}, 167 (1955).
\item{9.}
A. Brillet and J. L. Hall,  {\it Phys. Rev. Lett.} {\bf 42}, 549 (1979).
\item{10.}
D. C. Champeney, G. R. Isaak, and A. M. Khan, {\it
Phys. Lett.} {\bf 7}, 241 (1963).
\item{11.}
E. Riis, L.-U. A. Anderson, N. Bjerre, O. Poulsen,
S. A. Lee, and J. L. Hall,
{\it Phys. Rev. Lett.} {\bf 60}, 81 (1988).
\item{12.}
T. P. Krisher, L. Maleki, G. F. Lutes, L. E. Primas, R. T.
Logan, J. D. Anderson, and C. M. Will,
{\it Phys. Rev. D} {\bf 42}, R731 (1990).
\item{13.}
V. W. Hughes, H. G. Robinson, and V. Beltran-Lopez,
{\it Phys. Rev. Lett.} {\bf 4}, 342 (1960).
\item{14.}
R. W. P. Drever, {\it Phil. Mag.} {\bf 6}, 683 (1961).
\item{15.}
J. D. Prestage, J. J. Bollinger, W. M. Itano, and D. J. Wineland,
{\it Phys. Rev. Lett.} {\bf 54}, 2387 (1985).
\item{16.}
S. K. Lamoreaux, J. P. Jacobs, B. R. Heckel, F. J. Raab, and
E. N. Fortson, {\it Phys. Rev. Lett.} {\bf 57}, 3125 (1986).
\item{17.}
T. E. Chupp, R. J. Hoare, R. A. Loveman, E. R. Oteiza, J. M.
Richardson, M. E. Wagshul, and A. K. Thompson,
{\it Phys. Rev. Lett.} {\bf 63}, 1541 (1989).
\item{18.}
R. V. Pound and G. A. Rebka, {\it Phys. Rev. Lett.} {\bf 4}, 337
(1960).
\item{19.}
R. V. Pound and J. L. Snider {\it Phys. Rev.} {\bf 140}, B788 (1965).
\item{20.}
R. F. C. Vessot, M. W. Levine, E. M. Mattison, E. L. Blomberg,
T. E. Hoffman, G. U. Nystrom, B. F. Farrell, R. Decher,
P. B. Eby, C. R. Baugher, J. W. Watts, D. L. Teuber,
and F. D. Wills,
{\it Phys. Rev. Lett.} {\bf 45}, 2081 (1980).
\item{21.}
J. P. Turneaure, C. M. Will, B. F. Farrell, E. M. Mattison,
and R. F. C. Vessot, {\it Phys. Rev. D} {\bf 27}, 1705 (1983).
\item{22.}
A. Godone, C. Novero, and P. Tavella, {\it Phys. Rev. D} {\bf 51}, 319
(1995).
\item{23.}
J. D. Prestage, R. L. Tjoelker and L. Maleki, submitted to {\it Phys.
Rev. Lett.}
\item{24.}
A. I. Shlyakhter, {\it Nature} {\bf 264}, 340 (1976).
\item{25.}
P. G. Roll, R. Krotkov, and R. H. Dicke, {\it Ann.
Phys. (N.Y.)} {\bf 26}, 442 (1964).
\item{26.}
V. B. Braginsky and V. I. Panov, {\it Soviet Physics
JETP} {\bf 34}, 463 (1972) [{\it Zh. \'Eksp.  Teor. Fiz.} {\bf 61}, 873
(1971)].
\item{27.}
E. G. Adelberger, C. W. Stubbs, B. R.
Heckel, Y. Su, H. E. Swanson, G. Smith, J. Gundlach, and W. F. Rogers,
{\it Phys.  Rev. D} {\bf 42}, 3267 (1990).
\item{28.}
Y. Su, B. R. Heckel,
E. G. Adelberger, J. H.  Gundlach, M. Harris, G. L. Smith,
and H. E. Swanson,
{\it Phys.  Rev. D} {\bf 50}, 3614 (1994).
\item{29.}
C. M. Will and K. Nordtvedt, {\it Astrophys. J.} {\bf 177}, 757 (1972).
\item{30.}
T. Damour and K. Nordtvedt, {\it Phys.
Rev. Lett.} {\bf 70}, 2217 (1993).
\item{31.}
R. J. Hellings, P. J., Adams, J. D. Anderson, M. S. Keesey, E. L.
Lau, E. M. Standish, V. M. Canuto, and I. Goldman,
{\it Phys. Rev. Lett.} {\bf 51},
1609 (1983).
\item{32.}
I. I. Shapiro,
In {\it General Relativity and
Gravitation 1989}. edited by N. Ashby, D. F. Bartlett and W. Wyss
(Cambridge: Cambridge University Press, 1990), p. 313.
\item{33.}
V. M. Kaspi, J. H. Taylor, and M. F. Ryba, {\it Astrophys. J.} {\bf
428}, 713 (1994).
\item{34.}
R. A. Hulse and J. H. Taylor, {\it Astrophys. J. Lett.} {\bf 195}, L51
(1975).
\item{35.}
J. H. Taylor, In {\it General Relativity and Gravitation 1992}, edited
by R. J. Gleiser, C. N. Kozameh, and O. M. Moreschi (Bristol:
Institute of Physics, 1993), p. 287.
\item{36.}
T. Damour and J. H. Taylor, {\it Astrophys. J.} {\bf 366}, 501
(1991).
\item{37.}
A. Wolszczan, {\it Nature}, {\bf 350}, 688 (1991).
\item{38.}
T. A. Prince, S. B. Anderson, S. R. Kulkarni and A. Wolszczan,
{\it Astrophys. J. Lett.} {\bf 374}, L41 (1991).
\item{39.}
D. R. Stinebring, M. F. Ryba, J. H. Taylor, and R. W. Romani, {\it
Phys. Rev. Lett.} {\bf 65}, 285 (1990).
\item{40.}
A. Abramovici, W.E. Althouse, R.W.P. Drever,
Y. G{\" u}rsel, S. Kawamura, F.J. Raab, D. Shoemaker,
L. Siewers, R.E. Spero, K.S. Thorne, R.E. Vogt,
R. Weiss, S.E. Whitcomb, and M.E. Zucker,
{\it Science} {\bf 256}, 325 (1992).
\item{41.}
C. Cutler, T.A. Apostolatos, L. Bildsten, L.S.
Finn, E.E. Flanagan, D. Kennefick, D.M. Markovic,
A. Ori, E. Poisson, G.J. Sussman, and K.S.
Thorne, {\it Phys. Rev. Lett.} {\bf 70}, 2984 (1993).
\item{42.}
R. V. Wagoner and C. M. Will, {\it Astrophys. J.} {\bf 210}, 764
(1976).
\item{43.}
L. E. Kidder, C. M. Will, and A. G. Wiseman,
{\it Phys. Rev. D} {\bf 47}, R4183 (1993).
\item{44.}
L. Blanchet, T. Damour, B.R. Iyer, C.M.
Will, and A.G. Wiseman, {\it Phys. Rev. Lett.} (in press).
\item{45.}
H. Tagoshi and T. Nakamura, {\it Phys. Rev. D} {\bf 49}, 4016 (1994).
\item{46.}
H. Tagoshi and M. Sasaki, {\it Prog. Theor. Phys.} {\bf 92}, 745 (1994).
\item{47.}
N. Ashby, {\it GPS World}, Nov. 1993, p. 42.
\item{48.}
D. Kleppner, {\it Physics Today} {\bf 47}, 9 (Jan.) (1994).
\bigskip
\end